# Superconducting-Spin Reorientation in Spin-Triplet Multiple Superconducting Phases of UTe$_2$

Katsuki Kinjo[1*†], Hiroki Fujibayashi[1], Hiroki Matsumura[1], Fumiya Hori[1], Shunsaku Kitagawa[1], Kenji Ishida[1*], Yo Tokunaga[2], Hironori Sakai[2], Shinsaku Kambe[2], Ai Nakamura[3], Yusei Shimizu[3], Yoshiya Homma[3], Dexin Li[3], Fuminori Honda[3,4], and Dai Aoki[3,5]

[1]Department of Physics, Graduate School of Science, Kyoto University; Kyoto, Kyoto, 605-0953, Japan.

[2]Advanced Science Research Center, Japan Atomic Energy Agency; Tokai, Ibaraki, 319-1195, Japan.

[3]Institute for Materials Research, Tohoku University, Oarai, Ibaraki, 311-1313, Japan

[4]Central Institute of Radioisotope Science and Safety, Kyushu University, Fukuoka, 819-0395, Japan

[5]University Grenoble Alpes, CEA, Grenoble INP, IRIG, PHELIQS; F-38000 Grenoble, France

[†]Present address: Institute of Multidisciplinary Research for Advanced Materials, Tohoku University, Sendai, Miyagi, 980-8577, Japan

*Corresponding author. Email: katsuki.kinjo.c6@tohoku.ac.jp (K.K.); kishida@scphys.kyoto-u.ac.jp (K.I.).

**Abstract:** Superconducting (SC) state has spin and orbital degrees of freedom, and spin-triplet superconductivity shows multiple SC phases due to the presence of these degrees of freedom. However, the observation of spin-direction rotation occurring inside the SC state (SC spin rotation) has hardly been reported. UTe$_2$, a recently discovered topological superconductor, exhibits various SC phases under pressure: SC state at ambient pressure (SC1), high-temperature SC state above 0.5 GPa (SC2), and low-temperature SC state above 0.5 GPa (SC3). We performed nuclear magnetic resonance and AC susceptibility measurements on single-crystal UTe$_2$. The *b*-axis spin susceptibility remains unchanged in SC2, unlike in SC1, and decreases below the SC2-SC3 transition with spin modulation. These unique properties in SC3 arise from the coexistence of two SC order parameters. Our NMR results confirm the spin-triplet superconductivity with SC spin parallel to *b* in SC2, and unveil the remaining of spin degrees of freedom in superconducting UTe$_2$.

**One-Sentence Summary:** Multiple superconducting phases with spin rotation in the superconducting state was observed.

**Introduction**

What kind of an ordered state is realized under competing interaction is a central issue in condensed matter physics. Superconductivity and superfluidity are most remarkable macroscopic quantum phenomena produced by two-fermion pairs, so-called the Cooper pair. In conventional superconductivity, which covers almost all superconductors discovered so far, the Cooper pairs have zero total spin and orbital angular momenta, and have no degrees of freedom. Therefore, only one superconducting (SC) state is realized.

Theoretically, it is possible that either or both of the two angular momenta are non-zero, and hence the coexistence of two SC phases, and/or multiple SC phases was anticipated. The well-known example of such multi-phases is superfluid $^3$He, in which two phases (A and B) with lowest energy are realized in zero field, and becomes more complex under magnetic fields and/or anisotropic environment (*1-4*). The presence of the Majorana particles, which is applicable to "qubits" in quantum computer was suggested in the surface state of the B phase (*5*). However, in superconductors, there are only limited examples for such multiple SC phases, for example, UPt$_3$ (*6*), Th-doped UBe$_{13}$ (*7*), and CeRh$_2$As$_2$ (*8*) are known so far. In addition, SC-spin rotation, which is the smoking gun for the spin-triplet SC multiphase due to the presence of the spin degrees of freedom, has hardly been observed, because of the tiny difference between the two critical temperatures (*9*). It is quite important to find the textbook examples of the spin triplet superconductivity, corresponding to the superfluid $^3$He, and to find the phenomena related to the spin and orbital degrees of freedom.

Herein, we focus on a recently discovered uranium-based superconductor uranium ditelluride (UTe$_2$) with SC transition temperature $T_c \sim$ 1.6 K (*10*). UTe$_2$, crystallizing orthorhombic structure with space group *Immm* (#71, $D_{2h}$), as shown in Fig. 1A, has the multiple SC phases. Under pressure (*P*), the $T_c$ of UTe$_2$ increases to about 3 K at 1.2 GPa (*11-13*), as shown in Fig. 1B. Below 1.6 GPa, the two jumps in the temperature (*T*) dependence of specific heat indicate the existence of at least two SC phases (*11, 13*): one exists at ambient pressure (SC1) and its $T_c$ gradually decreases with applying *P*, and the other is an SC phase induced by *P* above 0.5 GPa (SC2), whose $T_c$ has a maximum at 1.2 GPa. As a result, $T_c$ of SC1 is below $T_c$ of SC2 above 0.5 GPa, and thus we call this SC state SC3 in this paper. Above 1.6 GPa, superconductivity suddenly disappears and a magnetic anomaly, which is considered as the antiferromagnetic state, was observed (*13*).

In addition, UTe$_2$ shows multiple SC phases even at ambient pressure. For $H \parallel b$, $T_c$ first decreases with increasing *H* and shows a minimum at around 16 T, then $T_c$ increases up to 35 T (*14*). The bulk SC properties in the high-field (HF) SC phase were recently confirmed by thermodynamic and $\chi_{AC}$ measurements (*15, 16*), suggesting that the pairing interaction grows up above 16 T (*15*).

Furthermore, recent nuclear magnetic resonance (NMR) measurements revealed a tiny change of spin susceptibility just below $T_c$ for $H \parallel b$ and $H \parallel c$ (*17-19*), and almost no change in the *a*-axis spin susceptibility (*20*). Above 14 T for $H \parallel b$, spin susceptibility remains constant below $T_c$ (*16, 18*). These results indicate that UTe$_2$ is a spin-triplet superconductor with the spin degrees of freedom and the SC order parameter **d**-vector, which is perpendicular to the SC spin, having the *b* and *c* components in the low-field (LF) region. In UTe$_2$, a complex interplay of various interactions was reported, such as Ising-like ferromagnetic fluctuations from NMR studies (*21*) and the antiferromagnetic fluctuations from the inelastic neutron scattering measurements (*22, 23*). Such competing interactions are thought to induce rich physics such as the HFSC phase in $H \parallel b$ and multiple SC phases under pressure. Thus, identifying the SC properties of the SC2 and SC3 states is the last piece to understanding the multiple SC phases of UTe$_2$.

To elucidate the SC properties of the multiple SC phases in UTe$_2$, nuclear magnetic resonance (NMR) serves as a highly effective tool. NMR is a technique that probes nuclear spins, which interact with electron spins through strong hyperfine interactions. Consequently, it enables us to investigate the electron spin susceptibility microscopically. While the spin susceptibility in the SC state cannot be measured due to the SC diamagnetism by the bulk magnetization measurement, NMR can detect it with the hyperfine interaction. Moreover, to measure spin susceptibility in the multiple SC states under pressure, NMR is a unique measurement that meets these requirements.

Here, we report the results of the NMR Knight shift $K$ for $H \parallel b$ ($K_b$) at 1.2 GPa, at which $T_c$ of SC2 is maximum. We found that the spin susceptibility along the $b$ axis is unchanged in SC2, which is different from that observed in SC1 at ambient pressure. More surprisingly, at the SC2-SC3 transition, $K_b$ suddenly decreases and the NMR-spectral width becomes broader. These results suggest that the SC order parameter of SC1 and SC2 is different and that SC2 is a spin-triplet superconductivity with spin oriented toward the $b$ axis and SC spin reoriented in SC3 ~~state~~. These results suggest that the SC order parameter $d$-vector was changed at the SC3 transition, and the spin degrees of freedom remains in the superconductivity of UTe$_2$.

**Results and Discussion**

As there are two crystallographically inequivalent Te sites in UTe$_2$, we observed two $^{125}$Te-NMR peaks as reported in the previous paper (*17−19*). An NMR peak with the smaller [larger] $K$ in $H \parallel b$ as Te(1) [Te(2)] by following the previous paper (*21*). The $T$ variation of the NMR spectrum of Te(2) at $P = 1.2$ GPa is shown in the Fig. 2A, and the $T$ dependence of the full width at half maximum (FWHM), the NMR Knight shift ($K$) determined with the NMR spectrum, and $\chi_{AC}$ are shown in Fig. 2B. Here, the NMR Knight shift is proportional to the spin susceptibility and the FWHM is proportional to the distribution of magnetization. As reported in the previous paper (*24-26*), $K_b$ exhibits a broad maximum at $T\chi_{max}$ and gradually decreases with decreasing $T$, similar to the $T$ dependence of $\chi$ (*24*). Owing to the increase in $T_c$ and the decrease in $T\chi_{max}$ with increasing $P$ (*24*), $K_b$ highly depends on $T$ around $T_c$ at 1.2 GPa, unlike the case of ambient pressure (0 GPa). To clarify the change in the Knight shift related to the SC transition, we calculate $\Delta K \equiv K_{SC}^{spin} - K_{normal}^{spin}$ as shown in Fig. 3A. Here, $K_{normal}^{spin}$ ($K_{SC}^{spin}$) is $K^{spin}$ in the normal (SC) state, and is estimated from the extrapolation above $T_c$. Although the absolute value of the spin part in $K$ ($K_{spin}$) decreases with decreasing $T$, $\Delta K$ is almost zero above 0.5 K, indicating that the spin susceptibility in SC2 is the same as that in the normal state for $H \parallel b$. This $T$ dependence of the spin susceptibility in SC2 is similar to that in the A phase of the $^3$He superfluid (*27*). Below 0.5 K, we observed a sudden decrease in $K_b$ and a broadening of the NMR spectrum, although no additional change was observed in $\chi_{AC}$. The magnitude of the decrease in $K_b$ below 0.5 K is almost the same as that in $K$ at 0 GPa (*17*). This is further evidence of the occurrence of the phase transition inside the SC state from the microscopic point of view.

Now, we consider possible SC order parameters suggested by the present NMR results. As discussed in the previous papers *(17-20)*, the order parameter of spin-triplet superconductivity is $d$-vector, which is perpendicular to the SC spin components. In other words, $K_i$ proportional to the spin susceptibility along the $i$ axis decreases when $d$-vector has the $i$ component, but is unchanged when the $d$-vector has no $i$ component. This corresponds to the SC spin pointing to the $i$ direction. The possible SC phases at zero field are listed in Table 1 of Ref. *(16)*. Our findings indicate that the $d$-vector has no $b$ component (the SC-spin is oriented to the $b$ axis) in SC2, and has a finite $b$ component in SC3, which is the same as that of SC1. This is the $d$-vector rotation, evidencing the spin-triplet multiple SC phases with spin-degrees of freedom.

Another important point of Fig. 2B is that the FWHM of the Te(2) site shows an additional increase below 0.5 K. Figures 3A, 3B, and 3C show the temperature dependence of the FWHM divided by the applied fields ($\mu_0 H$), which reflects the distribution of the spin-susceptibility, $\Delta K$ measured at the Te(2) site, and $\chi_{AC}$ measured at 0.8, 1.0, and 2.5 T, respectively. To clarify the increase in line width due to the normal-SC2 transition and the increase in line width due to the SC2-SC3 transition, we defined the values $\Delta\text{FWHM}^{\text{Meissner}}$ and $\Delta\text{FWHM}^{\text{SC-SC}}$. $\Delta\text{FWHM}^{\text{Meissner}}$ and $\Delta\text{FWHM}^{\text{SC-SC}}$ are defined as FWHM(SC2) - FWHM(Normal) and FWHM(SC3) - FWHM(SC2), respectively, as shown in Fig. 3D. In general, the FWHM of the NMR spectrum increases below $T_c$ due to SC diamagnetism. This effect is related to the SC penetration depth and was observed below 3.0 K in SC2, and was suppressed with increasing $\mu_0 H$, as shown in Fig. 3E. Decrease of $\Delta\text{FWHM}^{\text{Meissner}}/\mu_0 H$ with applying $H$ is interpreted by the conventional SC diamagnetic effect. However, the additional increase in the FWHM below 0.5 K ($\Delta\text{FWHM}^{\text{SC-SC}}/\mu_0 H$) cannot be explained by such SC diamagnetic effect, since $\chi_{AC}$ shows no anomaly around 0.5 K, and $\Delta\text{FWHM}^{\text{SC-SC}}/\mu_0 H$ is independent of the applied field, as shown in Fig. 3F. In addition, from the comparison of the NMR spectra between 0.6 K (SC2) and 0.13 K (SC3), it is noted that the spectrum shows a tail to the larger Knight-shift side than the Knight shift in SC2 as shown in Fig. 4A. This is quite unusual for the NMR spectrum in the SC state, as the spin susceptibility decreases in the SC state. Such an NMR spectrum variation in the SC state was not observed in SC1 [Fig. 4 (A)], although the spectrum broadening due to the SC diamagnetic effect was observed. The ratio between the FWHM of the Te(1) and Te(2) spectra in SC3 is almost unchanged to that in the normal state *(28)*. These results indicate that the additional linewidth broadening in SC3 arises from the unusual inhomogeneity of the spin susceptibility, which was not observed in SC1. Thus, SC3 is a different SC state from SC1.

We consider the several possibilities of the origin of linewidth broadening in SC3. One is due to the SC vortex state in SC3. Since spin susceptibility is unchanged in SC2, the NMR signal from the vortex core is observed at almost the same position as the SC NMR signal outside the vortex. On the other hand, the spin susceptibility decreases in SC3, and the signal from the vortex core should be observed at the different position from the SC NMR signal, resulting in the NMR signal becoming broader in the mixed state of SC3. However, this possibility would be unlikely, because the measurement fields are far below $H_{c2}^{\text{orb}} \sim 70$ T expected from Werthamer, Helfand, and Hohenberg (WHH) theory *(29)*. Thus, the area fraction of vortex is approximately 1%, and the contribution to the NMR spectrum is negligibly small. Another possibility is the spatial inhomogeneity with coexistence of the two SC states in the sample. However, this possibility is ruled out, as the observed spectra could not be explained by the simple coexistence of the SC2 and SC3 states *(28)*. These results indicate that, whatever the inhomogeneities in the sample, the NMR linewidth broadening intrinsically occurs below the SC2-SC3 transition, and is proportional to the magnetic field. In other words, it represents an increase in the distribution of the magnetic susceptibility in the SC3 state. At present we have no clear interpretation of this phenomenon, but we believe that we observed an effect due to the mixing of the two SC order parameters as in the case of (U,Th)Be$_{13}$ *(30, 31)*. Actually, it was pointed out that the nonunitary spin triplet superconductivity leads to a spin-density wave type modulation of $d$-vector from the theoretical studies *(32)*.

Figures 5A and 5B shows the schematic image of the $T$ vs $H$ along the $b$ axis ($H_b$) phase diagram of UTe$_2$ at ambient pressure (A) and 1.2 GPa (B), respectively (*13, 15, 18, 26*). At ambient pressure, $T_c(H)$ decreases in the LFSC (Low field superconducting) state, and increases in the HFSC (High field superconducting) state up to the metamagnetic field 35 T ($H_m$). In the LFSC state, $K_b$ slightly decreases below $T_c$ (*17*) (see Fig. 5B). As mentioned above, $K_b$ is unchanged in the HFSC state (*16*) as shown in Fig. 5C, indicating that the SC spin align to the $b$ axis. At the pressure ~ 1.2 GPa where $T_c$ becomes maximum, $H_m$ is suppressed to almost half of $H_m$ at $P = 0$ and there is no more L-shape behavior in $H_b$-$T$ phase diagram (*26*). Considering the similarity in spin susceptibility in the SC state between the HFSC and SC2 states and first order transition like disappearance of SC state and Kondo coherent state with applying field or pressure (*25, 33, 34*), we conclude that the HFSC evolves into SC2 with applying pressure. In addition, the SC1 phase hides under the SC2 phase and becomes the SC3 phase (coexisting phase of SC1 and SC2), while the SC3 phase appears only inside the SC2 phase.

Another similarity between HFSC and SC2 is that $T_c$ increases with increasing $K_b$; Figure 6 shows the values of $K_b$ under magnetic field or under pressure. When a magnetic field is applied, the zero-field extrapolated $T_c$ continues to increase with the application of the field (*34*), and $K_b$ also increases. This tendency was also observed in the pressure experiment. When $K_b$ is above 6.0 %, indicating the sample in SC2, $T_c$ increases up to 3 K with increasing $K_b$. The $T_c$ vs $K_b$ curves show similarity between under pressure and under magnetic field. It strongly suggests that HFSC and SC2 are the similar phase and have the same origin. Based on this scenario, we suggest that the spin susceptibility along the $b$ axis ($\chi_b$) is an important parameter for HFSC and SC2, which determine the $T_c$ of these states.

In conclusion, we have performed $^{125}$Te-NMR measurement on UTe$_2$ under pressure to investigate the SC properties and determine the SC order parameter of SC2 and SC3 phases. The $b$-axis spin susceptibility in SC2 has the same value as that of the normal state below 3.0 K above 0.5 K. Below 0.5 K, spin susceptibility decreases, and the NMR linewidth increases. These results indicate that the SC order parameter is changed, *i.e.*, SC spin rotation occurs at the SC2-SC3 transition, and the novel SC state with two SC order parameters in SC3. This is microscopic evidence for the remaining the spin-degrees of freedom in the superconductivity of UTe$_2$, inherent to the spin-triplet superconductivity.

**Materials and Methods**
**Sample Preparation**

A high-quality single-crystal UTe$_2$ sample was grown by the chemical vapor transport method with Iodine as a transport agent, details of which are described in ref. (*10, 36*). The single crystal of UTe$_2$ was an almost rectangular shape with $2 \times 1 \times 1$ mm$^3$, with 2 mm along the *a* axis. To improve $^{125}$Te-NMR signal intensity, the sample was synthesized with the $^{125}$Te-enriched (99.9 %) metal as discussed in ref. (*18*). The transition temperature ($T_c$) of this sample is 1.6 K at ambient pressure.

**AC susceptibility measurements**

In order to confirm $T_c$, we measured the high-frequency AC susceptibility $\chi_{AC}$ using the NMR tank circuit. In the superconducting state, the impedance of the circuit changes due to the Meissner effect, and thus the tuning frequency of the circuit drastically changes just below $T_c$.

**Applying pressure and pressure estimation**

The hydrostatic pressure is applied using a piston-cylinder-type cell made of NiCrAl and CuBe alloys as described in ref. (*37*). Daphne 7373 was used as a pressure medium. The applied pressure was estimated by the SC transition temperature of Pb with the formula of
$P = (7.181 − T_c(P))/0.364$ (*38*).

**Field Alignment**

We used a split-pair magnet to apply horizontal fields to the sample and a $^3$He-$^4$He dilution refrigerator that can be rotated about the vertical axis. With this setup, we could rotate the magnetic field direction within the *bc* plane of the sample. The direction of the magnetic field was confirmed by measuring the Te1 signal as described in ref. (*24*).

**NMR measurements**

An NMR spectrometer with a 100 W (at 0 dB input) power amplifier (Thamway Product: N146- 5049A) was used for the measurements. A conventional spin-echo method was used for NMR measurements in the temperature range from 0.1 to 4.2 K and in the magnetic field range from 0.8 to 2.5 T. All measurements were carried out with the $^3$He-$^4$He dilution refrigerator, in which the pressure cell was immersed into the $^3$He-$^4$He mixture to reduce radio frequency heating during measurements. The NMR spectrum was obtained by summation of the FFT spectrum with a 5 kHz step. The applied field is estimated by $^{65}$Cu-NMR measurements as discussed here (*39*).

**Acknowledgments:**

The authors thank A. Miyake, Y. Yanase, Y. Matsuda, K. Machida, S. Fujimoto, V. P. Mineev, Y. Maeno, S. Yonezawa, J-P. Brison, G., Knebel, W. Knafo for their valuable discussions. This work was supported by the Kyoto University LTM Center.

**Funding:**

This work was supported by Grants-in-Aid for Scientific Research (Grant Nos.JP19K03726, JP19K14657, JP19H04696, JP19H00646, JP20H00130, JP20KK0061, JP21K18600, JP22H04933, JP22H01168, and JP23H01124).

This work was also supported by JST SPRING (Grant Number JPMJSP2110).

KK, HF, and HM would like to acknowledge the support from the Motizuki Fund of Yukawa Memorial Foundation.

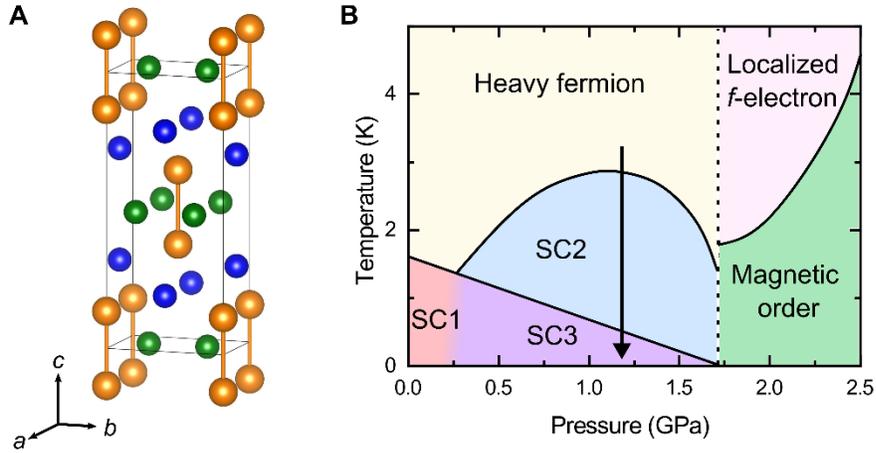

**Figure 1. Crystal structure and pressure-temperature phase diagram of UTe$_2$.**
(**A**) Crystal structure of UTe$_2$ made by VESTA (*35*). The U atom forms the dumbbell structure, and the dumbbells form a line along the *a* axis. (**B**) Pressure–temperature phase diagram of UTe$_2$. Bold lines correspond to the thermodynamic phase transitions detected by the specific-heat measurements (*11, 13*). The phase boundary between SC1 and SC3 has not been detected. The arrow represents the measurement range in this report.

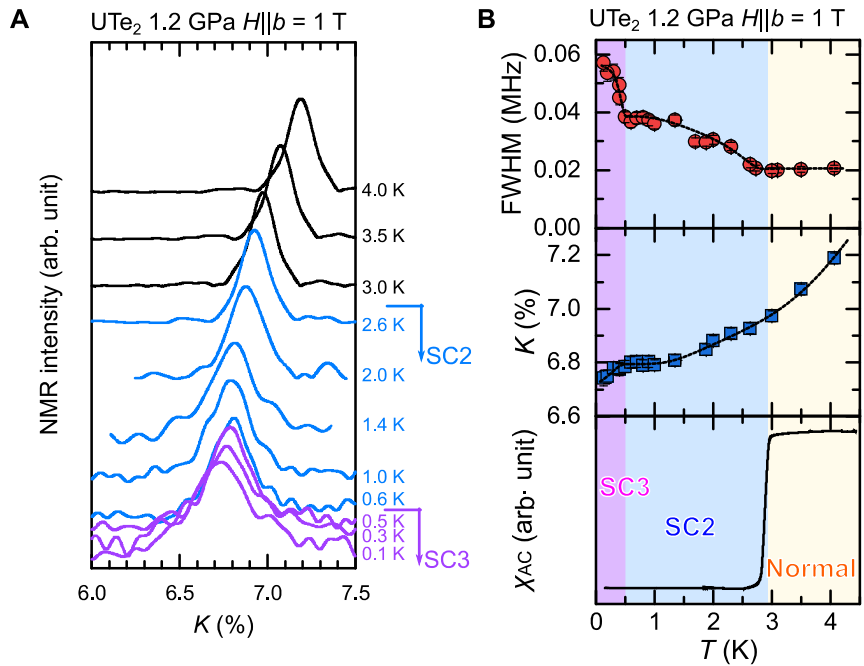

**Figure 2. NMR evidence for superconducting-spin rotation.**
(**A**) Temperature variation of the NMR spectra of UTe$_2$ at 1.2 GPa. (**B**) Temperature dependence of the full-width half-maximum (FWHM), peak position $K$ of Te2 signal, and $\chi_{AC}$ at 1.2 GPa. Line colors in panel (A) and background color in panel (B) represent the phase appearing in Fig. 1(B). The kinks in the T dependence of K and FWHM suggest the phase transition at around 0.5 K, although $\chi_{AC}$ does not show any anomaly. The dashed lines in top and middle panels are guides to eye.

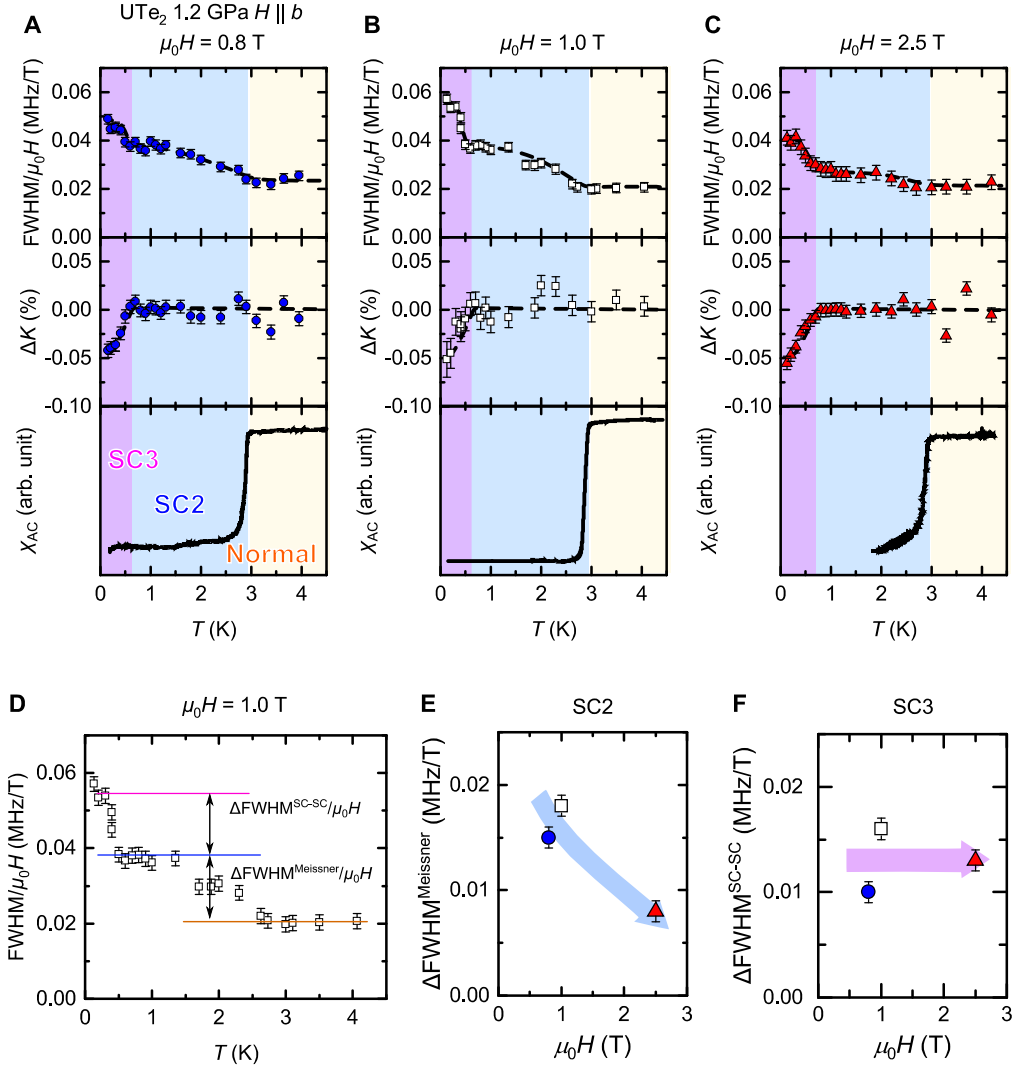

**Figure 3. NMR data evidencing the increase in magnetic inhomogeneity.**
Temperature dependence of (Top) the FWHM/$\mu_0H$, (Middle) $\Delta K$, and $\chi_{AC}$ at (**A**) 0.8 T, (**B**) 1.0 T, (**C**) 2.5 T of the Te2 NMR spectrum. (**D**) The definition of the $\Delta$FWHM$^{\text{Meissner}}$/$\mu_0H$ and $\Delta$FWHM$^{\text{Meissner}}$/$\mu_0H$, which are the NMR linewidth broadening in SC2 and SC3 respectively, is shown. The magnetic field dependence of (**E**) the $\Delta$FWHM$^{\text{Meissner}}$/$\mu_0H$ and (**F**) the $\Delta$FWHM$^{\text{SC-SC}}$/$\mu_0H$. The definitions of these two are given in the main text. $\Delta$FWHM$^{\text{Meissner}}$/$\mu_0H$ decreases with applying field. This is conventional superconducting behavior, and evidences the bulk superconductivity in the SC2 state. $\Delta$FWHM$^{\text{SC-SC}}$/$\mu_0H$ is almost constant against the field.

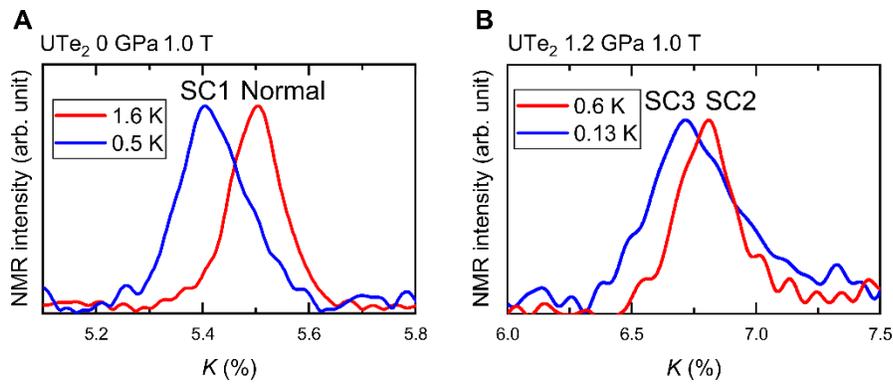

**Figure 4. Comparison of the NMR spectrum broadening in SC1 and SC3.**
**(A)** The NMR-spectrum variation occurring in SC1. **(B)** The NMR-spectrum broadening occurring in SC3. The spectrum in SC3 shows a tail to the larger Knight-shift side than the Knight shift in SC2.

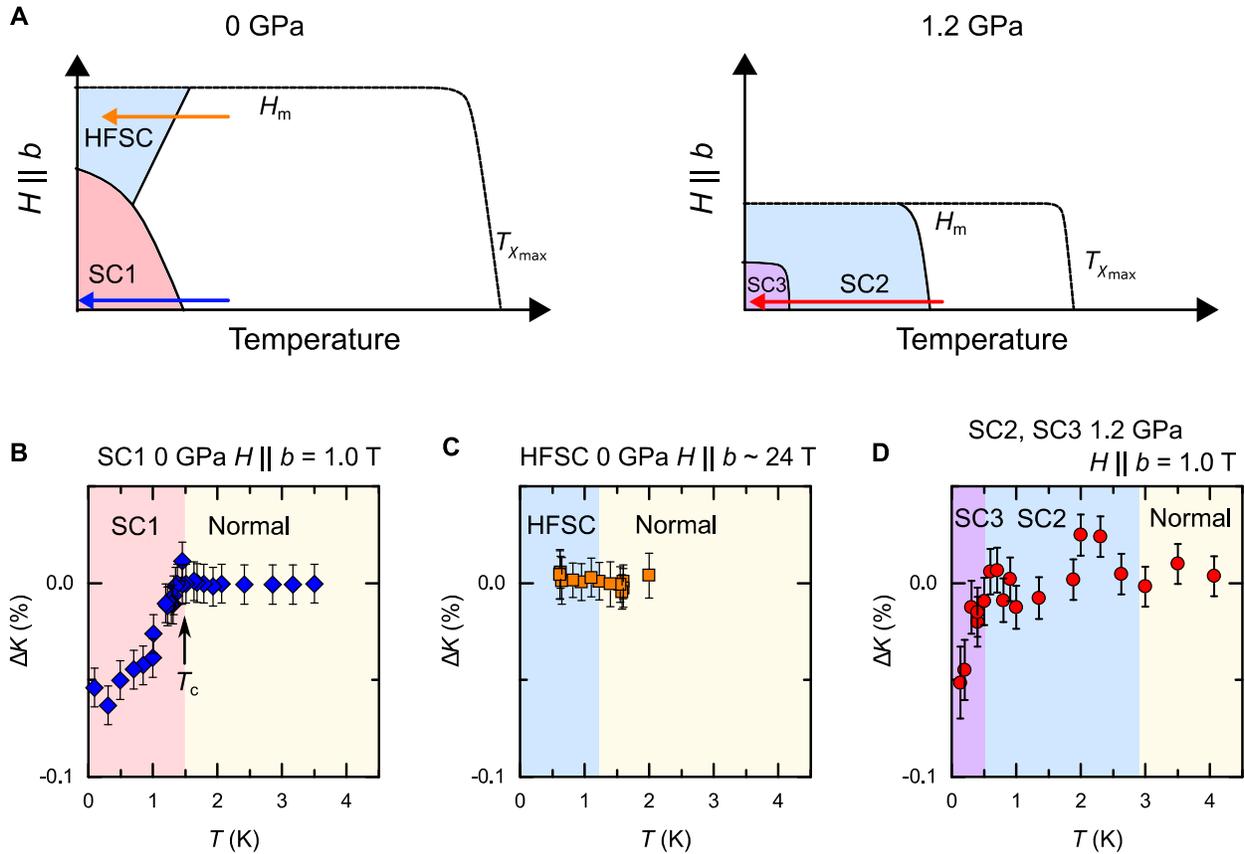

**Figure 5. General phase diagram expected from our NMR results.**
(**A**) The schematic images of the *H-T* phase diagram of UTe$_2$ at 0 GPa and 1.2 GPa. The bold lines are the transition lines determined by the thermodynamic measurements (*11, 13, 15*). The dashed line represents the temperature where the $\chi_b$ shows maximum ($T_{\chi\text{max}}$) and metamagnetic transition. (**B**)(**C**)(**D**) The temperature dependence of $\Delta K$ in 4 states appears in panel (**A**). The colored arrows in panel (**A**) represent the measurement range in panels (**B,C,D**).

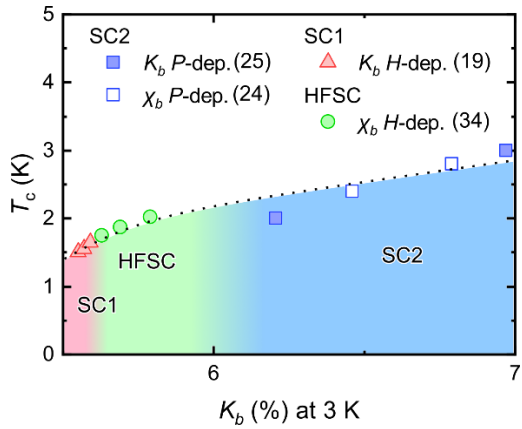

**Figure 6. Relation between *b*-axis susceptibility and critical temperature**
This diagram shows relation between *b*-axis magnetic susceptibility and the $T_c$ of superconducting phases. For $K_b$ between 5.7 and 6.0, $T_c$ is affected by the magnetic field along the *b*-axis. The extrapolation of $T_c$ to $H = 0$ in HFSC is based on the strong coupling parameter of Miyake *et al* (*34*). The magnetic field enhances $K_b$ and increases the $T_c$. As $K_b$ increases further, $T_c$ enters the SC2 region, and when $K_b$ increases to nearly 7%, $T_c$ increases to a maximum of 3 K. The dotted line is guides to eyes.